\documentclass{article}

\divide \topmargin by 2 \advance \topmargin -2in
\divide \leftmargin by 2 \advance \leftmargin -1in
\oddsidemargin \leftmargin \evensidemargin \leftmargin

\usepackage{amsthm,amsmath}
\usepackage{graphicx}
\usepackage{booktabs,multirow,colortbl}
\RequirePackage{hyperref}
\usepackage[utf8]{inputenc} 

\usepackage{pseudocode}
\newcounter{algccc}
\newcommand{\ccc}{\addtocounter{algccc}{1}\thealgccc.\ }

\newcommand{\qqa}{\hspace*{10pt}}
\newcommand{\qqb}{\qqa\qqa}
\newcommand{\qqc}{\qqb\qqa}
\newcommand{\qqd}{\qqc\qqa}
\newcommand{\qqe}{\qqd\qqa}
\newcommand{\qqf}{\qqe\qqa}

\def\pp{\mathinner{\ldotp\ldotp}}

\begin{document}

\title{SPRINT: Ultrafast protein-protein interaction prediction of the entire human interactome}

\author{\textsc{Yiwei Li} and  \textsc{Lucian Ilie}\footnote{Corresponding author: \texttt{ilie@uwo.ca}}\\
\ \\
Department of Computer Science\\ 
The University of Western Ontario\\
London, ON, Canada, N6A 5B7}

\maketitle

\begin{abstract}
\noindent
\textbf{Background:} Proteins perform their functions usually by interacting with other proteins. Predicting which proteins interact is a fundamental problem. Experimental methods are slow, expensive, and have a high rate of error. Many computational methods have been proposed among which sequence-based ones are very promising. However, so far no such method is able to predict effectively the entire human interactome: they require too much time or memory.\\
\textbf{Results:} We present SPRINT (Scoring PRotein INTeractions), a new sequence-based algorithm and tool for predicting protein-protein interactions. We comprehensively compare SPRINT with state-of-the-art programs on seven most reliable human PPI datasets and show that it is more accurate while running orders of magnitude faster and using very little memory. SPRINT is the only program that can predict the entire human interactome. Our goal is to transform the very challenging problem of predicting the entire human interactome into a routine task. \\
\textbf{Availability:} The source code of SPRINT is freely available from 
 \texttt{github.com/lucian-ilie/SPRINT/} and the datasets and predicted PPIs from \texttt{www.csd.uwo.ca/faculty/ilie/SPRINT/}.
\end{abstract}

\section{Background}

Protein-protein interactions (PPI) play a key role in many cellular processes since proteins usually perform their functions by interacting with other proteins. Genome-wide identification of PPIs is of fundamental importance in understanding the cell regulatory mechanisms \cite{Bonetta10_interactome} and PPI identification is one of the major objectives of systems biology. Various experimental techniques for identifying PPIs have been developed, most notably high throughput procedures such as two-hybrid assay and affinity systems \cite{Shoemaker07_PPIexperim}. Such methods are slow and expensive and have a high rate of error. A variety of computational methods have been designed to help predicting PPIs, employing sequence homology, gene co-expression, phylogenetic profiles, etc. \cite{Shoemaker07_PPIcompute,Liu12_survey,Zahiri13_survey}.

Sequence-based approaches \cite{Martin05_PPIpred,Pitre06_PIPE,Shen07_PPIpred,Guo08_PPIpred,HampRost15_PPI-PK} are faster and cheaper and can be used in addition to other methods, to improve their performance. Several top methods were evaluated by Park~\cite{Park09_PPIeval}. Park and Marcotte~\cite{ParkMarcotte12_C123} made the crucial observation that the datasets previously used for evaluation were biased due to the frequent occurrence of protein pairs common to testing and training data. They have shown that the prediction of the algorithms on the testing protein pairs is improved when the protein sequences are seen in training. To avoid this bias, they have built datasets of three levels of difficulty such that the predictive performance on these datasets generalizes to the population level. 
The performance of the top methods tested by Park~\cite{Park09_PPIeval} on the unbiased datasets of \cite{ParkMarcotte12_C123} was significantly lower than previously published, thus raising the bar higher for sequence-based methods. 

We introduce a new sequence-based PPI prediction method, SPRINT (Scoring PRotein INTeractions), that is more accurate than the current state-of-the-art methods as well as orders of magnitude faster.
The SPRINT algorithm relies on the same basic hypothesis that underlies most sequence-based approaches: a pair of proteins that are pairwise similar with a pair of interacting proteins has a higher chance to interact. However, the way this idea is used is very different. Similar regions are identified using an effective multiple spaced-seed approach and then processed to eliminate elements that occur too often to be involved in interactions. Finally, a score is computed for each protein pair such that high scores indicate increased probability of interactions. Details are given in the Methods section.

We compared SPRINT with the top programs considered by Park~\cite{Park09_PPIeval} and Park and Marcotte~\cite{ParkMarcotte12_C123} as well as the new method of Ding \textit{et al.}~\cite{Ding16_MMI}. The closest competitors are the machine learning-based programs of Ding \textit{et al.}~\cite{Ding16_MMI} and Martin \textit{et al.}~\cite{Martin05_PPIpred}, and PIPE2~\cite{Pitre06_PIPE,Pitre08_PIPE2}, which does not use machine learning. All comparisons are done using human datasets. 

To comprehensively compare the performance, we use multiple datasets, built according to the procedure of Park and Marcotte~\cite{ParkMarcotte12_C123} from six of the most reliable human PPI databases: Biogrid, HPRD, InnateDB (experimentally validated and manually curated PPIs), IntAct, and MINT. SPRINT provides the best predictions overall, especially for the more difficult C2 and C3 types.

Then, we used the entire human interactome to compare the speed. The comparisons of \cite{Park09_PPIeval} and \cite{ParkMarcotte12_C123} used fairly small datasets for comparison. In reality, these programs are meant to be used on entire proteomes and interactomes, where all protein sequences and known interactions are involved. SPRINT is several orders of magnitude faster. It takes between 15 and 100 minutes on a 12-core machine while the closest competitor, Ding's program, requires weeks and Martin's and PIPE2 require years. Moreover, Ding's program is unable to run the larger datasets as its memory requirements are very high.

The source code of SPRINT is freely available.


\section{Results}

We compare in this section SPRINT with several state-of-the-art sequence-based programs for PPI prediction on the most important human PPI datasets available. We focus on accurate prediction of the entire human interactome and therefore we have been using only human datasets. We start with a discussion concerning the datasets employed, as the way they are constructed can significantly impact the performance of the predicting programs.

\subsection{Park and Marcotte's evaluation scheme}

Park and Marcotte~\cite{ParkMarcotte12_C123} noticed that all methods have significantly higher performance for the protein pairs in the testing data whose sequences appear also in the training data. Three cases are possible, depending on whether both proteins in the test data appear in training (C1), only one appears (C2), or none (C3). They show that essentially all datasets previously used for cross validation are very close to the C1 type, whereas in the HIPPIE meta-database of human PPIs \cite{Schaefer12_HIPPIE} the C1-type human protein pairs accounts for only 19.2\% of these cases, whereas C2-type and C3-type pairs make up 49.2\% and 31.6\%, respectively. Therefore, testing performed on C1-type data is not expected to generalize well to the full population. The authors proceeded to designing three separate human PPI datasets that follow the C1, C2, and C3-type rules. 

A variation of the C1-3 idea was proposed by Hamp and Rost~\cite{HampRost15_newC1-3} where presence or absence of a protein is replaced by presence or absence of a similar protein. To avoid confusion, denote the new types C$'$1-3. That is, C$'$3 contains only protein pairs for which no similar ones are included in the training data. First, the construction of these datasets requires the introduction of an arbitrary similarity parameter. Second, Hamp and Rost \cite{HampRost15_newC1-3} did not evaluate the relevance of the new C$'$1-3 for full population generalization. Our computations show that the relevance of the C$'$2 and, especially, C$'$3 tests decreases significantly. We have computed that C$'$2-type pairs now amount for 48.3\% whereas C$'$3-type pairs make up only 16.5\% of the pairs in the human data. Therefore, the C$'$1-3 datasets are less relevant than C1-3 for full population generalization. For these reasons, we decided to use the C1-3 procedure of Park and Marcotte~\cite{ParkMarcotte12_C123}.

\subsection{Datasets}

We first describe the procedure of Park and Marcotte~\cite{ParkMarcotte12_C123} in detail. The protein sequences are from UniProt~\cite{Uniprot11}. The interactions were downloaded from the protein interaction network analysis platform \cite{Wu09_PINA} that integrates data from six public PPI databases: IntAct \cite{Kerrien07_IntAct}, MINT \cite{Chatr07_MINT}, BioGRID \cite{Stark11_BioGRID}, DIP \cite{Salwinski04_DIP}, HPRD \cite{Prasad09_HPRD} and MIPS MPact \cite{Guldener06_MIPS_MPact}. The datasets were processed by \cite{ParkMarcotte12_C123} as follows. Proteins in each data set were clustered using CD-HIT2~\cite{Li06_cd-hit} such that they shared sequence identity less than 40\%. Proteins with less than 50 amino acids as well as homo-dimeric interactions were removed. Negative PPI data were generated by randomly sampling protein pairs that are not known to interact. See \cite{ParkMarcotte12_C123} for more details.

The total number of proteins used is 20,117, involving 24,718 PPIs. The training and testing datasets are divided into forty splits (from the file human\_random.tar.gz), each consisting of one training file and three testing files, one for each type C1, C2, C3.  Therefore, each C1, C2, or C3 curve produced is the average of forty curves. In addition, they tested also 40-fold cross validation on the entire PPI set. In reality, the ratio between interacting and noninteracting protein pairs is believed to be 1:100 or lower. However, this would make it very slow or impossible to run some of the algorithms. Therefore, Park and Marcotte decided to use ratio 1:1.

We have used Park and Marcotte's procedure to design similar testing datasets using six other human PPI databases. Among the most widely known human PPI databases we have chosen six that appear to be the most widely used: Biogrid, HPRD, InnateDB (experimentally validated and manually curated PPIs), IntAct, and MINT. We have used 20,160 human protein sequences downloaded from UniProt. The protein sequences and interactions were downloaded in Oct.~2016. We perform four tests for each program on each dataset: 10 fold cross-validation using all PPIs and C1, C2, and C3 tests, the datasets for which are built as explained above, with the ratio between training and testing pairs of 10:1. The details of all datasets are given in Table~\ref{table_datasets}. 

\begin{table}[h]
\centering
\caption{The datasets used for comparing PPI prediction methods. The second column contains the total number of PPIs, while the third the fourth columns give the number of PPIs used for training and testing, respectively, in the C1, C2, and C3 tests.}
\label{table_datasets}
\begin{tabular}{@{}lrrrl@{}}
\toprule
Dataset & \multicolumn{3}{c}{PPIs} & Website  \\ \cmidrule{2-4}
	 & All & Training & Testing &  \\ \midrule
Park and Marcotte & 24,718 & 14,186 & 1,250 & \texttt{www.marcottelab.org/differentialGeneralization} \\ 
Biogrid  & 215,029 & 100,000 & 10,000 & \texttt{thebiogrid.org} \\
HPRD Release 9 & 34,044 & 10,000 & 1,000 & \texttt{www.hprd.org} \\ 
InnateDB experim.~validated & 165,655 & 65,000 & 6,500 & \texttt{www.innatedb.com} \\ 
InnateDB manually curated & 9,913 &3,600 & 360 & \texttt{www.innatedb.com} \\ 
IntAct & 111,744 & 52,500 & 5,250 & \texttt{www.ebi.ac.uk/intact} \\ 
MINT & 16,914 & 7,000 & 700 & \texttt{mint.bio.uniroma2.it} \\ \bottomrule
\end{tabular}
\end{table}

\subsection{Competing methods}

We have compared SPRINT with the four methods considered by \cite{ParkMarcotte12_C123}. Three of those use machine learning: \cite{Martin05_PPIpred}, \cite{Shen07_PPIpred}, and \cite{Guo08_PPIpred}, whereas the fourth does not: PIPE~\cite{Pitre06_PIPE}. Since the first three methods do not have names, we use the first author's name to identify them: Martin \cite{Martin05_PPIpred}, Shen \cite{Shen07_PPIpred}, and Guo \cite{Guo08_PPIpred}. Note that we have tested the improved PIPE2~\cite{Pitre08_PIPE2}, the same version that was tested by Park and Marcotte. 

Many programs have been proposed for PPI prediction \cite{chang2010predicting,zhang2011adaptive,Zahiri13_PPIevo,zhang2014prediction,Zahiri14LocFuse,you2015predicting,you2017improved}, however, very few are available. We have obtained the source code for two programs: the PPI-PK method of \cite{HampRost15_PPI-PK} and the program of Ding \textit{et al.}~\cite{Ding16_MMI}.

Unfortunately, the PPI-PK method is too slow to be tested. For example, on the toy dataset of \cite{HampRost15_PPI-PK} that contains 200 proteins, 150 training pairs and 150 testing pairs, Martin's method took 0.01s to train whereas PPI-PK took 134s to train. The low speed makes it impossible to test on any meaningful datasets. For example, Martin's program completed training on a single test (out of forty) of the yeast dataset C1 of \cite{HampRost15_PPI-PK} in 18s whereas PPI-PK did not finish in 13 days. Running times were not given by \cite{HampRost15_PPI-PK}. We contacted the first author of PPI-PK~\cite{HampRost15_PPI-PK} and he confirmed that the parallel version of the program is not currently working.

We managed to run the program of Ding \textit{et al.}~\cite{Ding16_MMI} on all datasets. After eliminating the programs of Shen \textit{et al.}~\cite{Shen07_PPIpred} and Guo \textit{et al.}~\cite{Guo08_PPIpred} as placing last on the first datasets, comparison on all subsequent tests were performed against Martin, PIPE2, and Ding. 

Note that PIPE2 and SPRINT do not require negative training data as they do not use machine learning algorithms. All the other programs require both positive and negative training sets. Note also that Ding's program uses also additional information concerning electrostatic and hydrophobic properties of amino acids.

\subsection{Performance comparison}

\subsubsection{Park and Marcotte datasets}
\begin{figure}
\centering
\includegraphics{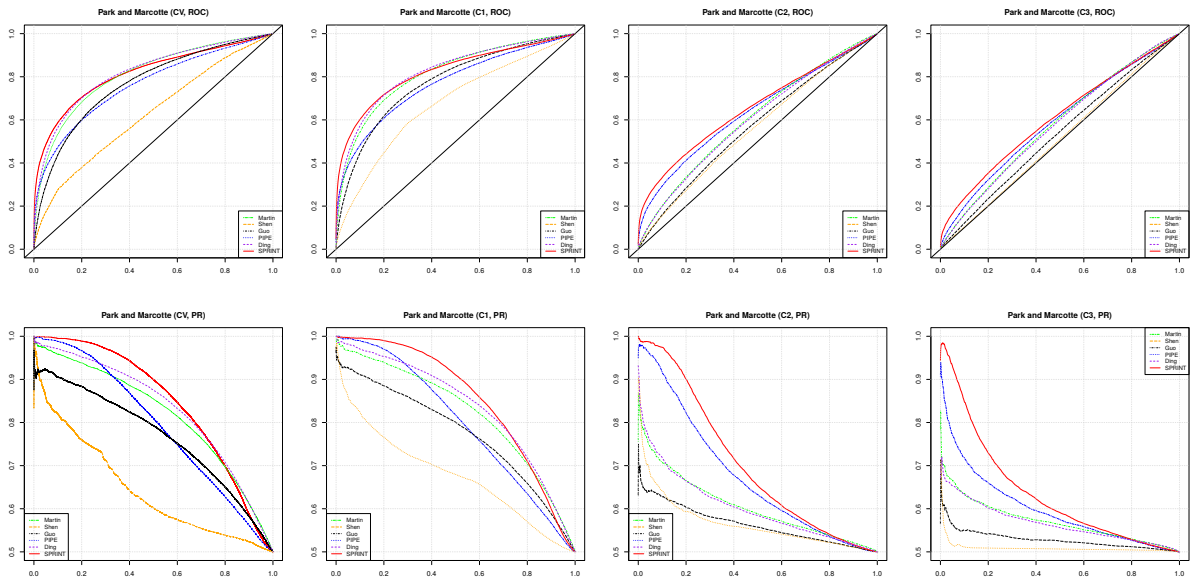}
  \caption{\textbf{Performance comparison on Park and Marcotte datasets.} The ROC curves (top row) and PR curves (bottom row) for CV, C1, C2, and C3 datasets, from left to right. \label{fig_Park_Marcotte_curves}}
\end{figure}

We present first the comparison of all five methods considered on the datasets of Park and Marcotte in Figure~\ref{fig_Park_Marcotte_curves}. The receiver operating characteristic (ROC) and precision-recall (PR) curves for the four tests, CV, C1, C2, and C3, are presented. 

The prediction performance on CV and C1 is very similar. The performance decreases from C1 to C2 and again to C3, both for ROC and PR curves. This is expected due to the way the datasets are constructed. The ROC curves do not distinguish very well between the prediction performance of the five methods. The difference is more clear in the PR curves. The SPRINT curve is almost always on top, especially at the beginning of the curve, where it matters the most for prediction. Ding's and Martin's are very close for CV and C1 datasets, followed by PIPE2. For C2 and C3 tests, the performance of Ding's and Martin's programs deteriorates and PIPE2 advances in second position.

\subsubsection{Seven human PPI databases}

For a comprehensive comparison, we have compared the top four programs on six datasets, computed as mentioned above from six databases: Biogrid, HPRD Release 9, InnateDB (experimentally validated and manually curated PPIs), IntAct, and MINT. Since the prediction on the CV datasets is similar with C1, we use only C1, C2 and C3 datasets.

For the purpose of predicting new PPIs, the behaviour at high specificity is important. We therefore compare the sensitivity, precision and $F_1$-score for several high specificity values. The table with all values is given in the supplementary material. We present here in Table~\ref{table_cutoff_avg} the average values for each dataset type (C1, C2, and C3) over all datasets for each specificity value. At the bottom of the table we give also the average over all three dataset types. The performance of SPRINT with respect to all three measures, sensitivity, precision, and $F_1$-score is the highest. Only Ding comes close for C1 datasets. the overall average of SPRINT is much higher than Ding's. PIPE2 comes third and Martin last. The performance of PIPE2 decreases much less from C1 to C3 compared with Ding's. It should be noted that a weighted overall average, where the contribution of each dataset type C1,2,3 is proportional with its share of the general population, would place PIPE2 slightly ahead of Ding.

The area under the ROC and PR curves is given in Table~\ref{table_AUC} for all seven datasets, including the C1-, C2-, and C3-average, as well as the overall average across types. Ding is the winner for the C1 tests and SPRINT is the winner for the C2 and C3 tests. In the overall average, SPRINT comes on top. Martin is third and PIPE2 last.

All ROC and PR curves are included in the supplementary material.

\begin{table}[h] 
\footnotesize
  \centering
\caption{Performance comparison at high specificity. Sensitivity, precision, and $F_1$-score averages for seven datasets are given for each dataset type C1, C2 and C3, as well as overall averages across types. Darker colours represent better results. The best results are in bold. \label{table_cutoff_avg}} 
\hspace*{-50pt}    \begin{tabular}{|@{}c@{}|@{\ }c@{\ }|cccc|cccc|cccc|}
    \toprule
    \multirow{2}[4]{*}{\textbf{Dataset}} & \multirow{2}[4]{*}{\textbf{Specificity}} & \multicolumn{4}{c|}{\textbf{Sensitivity}} & \multicolumn{4}{c|}{\textbf{Precision}} & \multicolumn{4}{c|}{\textbf{F1-score}} \\
\cmidrule{3-14}          &       & \textbf{$\!\!\!$Martin$\!\!\!$} & \textbf{$\!\!\!$PIPE2$\!\!\!$} & \textbf{$\!\!\!$Ding$\!\!\!$} & \textbf{$\!\!\!$SPRINT$\!\!\!$} & \textbf{$\!\!\!$Martin$\!\!\!$} & \textbf{$\!\!\!$PIPE2$\!\!\!$} & \textbf{$\!\!\!$Ding$\!\!\!$} & \textbf{$\!\!\!$SPRINT$\!\!\!$} & \textbf{$\!\!\!$Martin$\!\!\!$} & \textbf{$\!\!\!$PIPE2$\!\!\!$} & \textbf{$\!\!\!$Ding$\!\!\!$} & \textbf{$\!\!\!$SPRINT$\!\!\!$} \\
    \midrule
    \multirow{5}[1]{*}{\begin{tabular}{c}\textbf{C1}\\ \textbf{average}\end{tabular}} & 99.95 & \cellcolor[rgb]{ .988,  .894,  .839} 6.07 & \cellcolor[rgb]{ .945,  .78,  .678} 7.60 & \cellcolor[rgb]{ .82,  .459,  .22} 11.93 & \cellcolor[rgb]{ .776,  .349,  .067} \textbf{13.35} & \cellcolor[rgb]{ .796,  .392,  .125} 98.52 & \cellcolor[rgb]{ .788,  .376,  .106} 98.82 & \cellcolor[rgb]{ .988,  .894,  .839} 88.05 & \cellcolor[rgb]{ .776,  .349,  .067} \textbf{99.37} & \cellcolor[rgb]{ .988,  .894,  .839} 11.06 & \cellcolor[rgb]{ .945,  .78,  .678} 13.55 & \cellcolor[rgb]{ .824,  .467,  .235} 20.39 & \cellcolor[rgb]{ .776,  .349,  .067} \textbf{22.93} \\
          & 99.90 & \cellcolor[rgb]{ .988,  .894,  .839} 6.53 & \cellcolor[rgb]{ .929,  .741,  .624} 9.20 & \cellcolor[rgb]{ .816,  .447,  .208} 14.24 & \cellcolor[rgb]{ .776,  .349,  .067} \textbf{15.91} & \cellcolor[rgb]{ .827,  .475,  .243} 97.36 & \cellcolor[rgb]{ .796,  .392,  .129} 98.61 & \cellcolor[rgb]{ .988,  .894,  .839} 90.65 & \cellcolor[rgb]{ .776,  .349,  .067} \textbf{99.29} & \cellcolor[rgb]{ .988,  .894,  .839} 11.88 & \cellcolor[rgb]{ .925,  .733,  .608} 16.45 & \cellcolor[rgb]{ .82,  .455,  .22} 24.10 & \cellcolor[rgb]{ .776,  .349,  .067} \textbf{27.03} \\
          & 99.50 & \cellcolor[rgb]{ .988,  .894,  .839} 17.27 & \cellcolor[rgb]{ .922,  .718,  .588} 21.41 & \cellcolor[rgb]{ .776,  .349,  .067} \textbf{29.90} & \cellcolor[rgb]{ .784,  .369,  .094} 29.50 & \cellcolor[rgb]{ .988,  .894,  .839} 96.66 & \cellcolor[rgb]{ .878,  .612,  .439} 97.52 & \cellcolor[rgb]{ .792,  .384,  .114} 98.20 & \cellcolor[rgb]{ .776,  .349,  .067} \textbf{98.30} & \cellcolor[rgb]{ .988,  .894,  .839} 28.62 & \cellcolor[rgb]{ .914,  .694,  .557} 34.73 & \cellcolor[rgb]{ .78,  .353,  .071} 45.19 & \cellcolor[rgb]{ .776,  .349,  .067} \textbf{45.22} \\
          & 99.00 & \cellcolor[rgb]{ .988,  .894,  .839} 25.48 & \cellcolor[rgb]{ .945,  .776,  .671} 28.73 & \cellcolor[rgb]{ .8,  .404,  .145} 38.72 & \cellcolor[rgb]{ .776,  .349,  .067} \textbf{40.14} & \cellcolor[rgb]{ .988,  .894,  .839} 95.55 & \cellcolor[rgb]{ .898,  .659,  .506} 96.40 & \cellcolor[rgb]{ .804,  .416,  .161} 97.28 & \cellcolor[rgb]{ .776,  .349,  .067} \textbf{97.52} & \cellcolor[rgb]{ .988,  .894,  .839} 39.14 & \cellcolor[rgb]{ .933,  .753,  .639} 43.69 & \cellcolor[rgb]{ .8,  .412,  .153} 54.69 & \cellcolor[rgb]{ .776,  .349,  .067} \textbf{56.58} \\
          & 95.00 & \cellcolor[rgb]{ .902,  .671,  .522} 55.35 & \cellcolor[rgb]{ .988,  .894,  .839} 48.07 & \cellcolor[rgb]{ .776,  .349,  .067} \textbf{65.68} & \cellcolor[rgb]{ .824,  .463,  .231} 62.02 & \cellcolor[rgb]{ .886,  .627,  .459} 91.44 & \cellcolor[rgb]{ .988,  .894,  .839} 90.19 & \cellcolor[rgb]{ .776,  .349,  .067} \textbf{92.72} & \cellcolor[rgb]{ .804,  .42,  .165} 92.41 & \cellcolor[rgb]{ .898,  .659,  .502} 68.37 & \cellcolor[rgb]{ .988,  .894,  .839} 62.09 & \cellcolor[rgb]{ .776,  .349,  .067} \textbf{76.41} & \cellcolor[rgb]{ .816,  .447,  .204} 73.90 \\
\cmidrule{1-1}
    \multirow{5}[0]{*}{\begin{tabular}{c}\textbf{C2}\\ \textbf{average}\end{tabular}} & 99.95 & \cellcolor[rgb]{ .988,  .894,  .839} 5.55 & \cellcolor[rgb]{ .922,  .722,  .592} 10.65 & \cellcolor[rgb]{ .941,  .769,  .663} 9.22 & \cellcolor[rgb]{ .776,  .349,  .067} \textbf{21.45} & \cellcolor[rgb]{ .988,  .894,  .839} 96.33 & \cellcolor[rgb]{ .792,  .384,  .114} 99.42 & \cellcolor[rgb]{ .886,  .631,  .467} 97.92 & \cellcolor[rgb]{ .776,  .349,  .067} \textbf{99.62} & \cellcolor[rgb]{ .988,  .894,  .839} 9.78 & \cellcolor[rgb]{ .906,  .682,  .541} 18.91 & \cellcolor[rgb]{ .945,  .78,  .678} 14.69 & \cellcolor[rgb]{ .776,  .349,  .067} \textbf{33.16} \\
          & 99.90 & \cellcolor[rgb]{ .988,  .894,  .839} 5.88 & \cellcolor[rgb]{ .925,  .729,  .604} 11.28 & \cellcolor[rgb]{ .941,  .776,  .671} 9.78 & \cellcolor[rgb]{ .776,  .349,  .067} \textbf{23.40} & \cellcolor[rgb]{ .988,  .894,  .839} 93.66 & \cellcolor[rgb]{ .792,  .388,  .122} 98.96 & \cellcolor[rgb]{ .898,  .659,  .506} 96.11 & \cellcolor[rgb]{ .776,  .349,  .067} \textbf{99.34} & \cellcolor[rgb]{ .988,  .894,  .839} 10.40 & \cellcolor[rgb]{ .91,  .694,  .553} 19.98 & \cellcolor[rgb]{ .945,  .784,  .682} 15.70 & \cellcolor[rgb]{ .776,  .349,  .067} \textbf{36.08} \\
          & 99.50 & \cellcolor[rgb]{ .988,  .894,  .839} 11.73 & \cellcolor[rgb]{ .91,  .694,  .553} 19.52 & \cellcolor[rgb]{ .941,  .769,  .663} 16.59 & \cellcolor[rgb]{ .776,  .349,  .067} \textbf{32.73} & \cellcolor[rgb]{ .988,  .894,  .839} 93.86 & \cellcolor[rgb]{ .831,  .49,  .267} 97.11 & \cellcolor[rgb]{ .976,  .867,  .796} 94.11 & \cellcolor[rgb]{ .776,  .349,  .067} \textbf{98.22} & \cellcolor[rgb]{ .988,  .894,  .839} 20.17 & \cellcolor[rgb]{ .902,  .667,  .514} 31.86 & \cellcolor[rgb]{ .941,  .769,  .663} 26.59 & \cellcolor[rgb]{ .776,  .349,  .067} \textbf{47.77} \\
          & 99.00 & \cellcolor[rgb]{ .988,  .894,  .839} 15.03 & \cellcolor[rgb]{ .898,  .659,  .502} 24.93 & \cellcolor[rgb]{ .922,  .714,  .584} 22.55 & \cellcolor[rgb]{ .776,  .349,  .067} \textbf{37.60} & \cellcolor[rgb]{ .988,  .894,  .839} 91.85 & \cellcolor[rgb]{ .835,  .498,  .278} 95.64 & \cellcolor[rgb]{ .922,  .722,  .592} 93.52 & \cellcolor[rgb]{ .776,  .349,  .067} \textbf{97.07} & \cellcolor[rgb]{ .988,  .894,  .839} 25.26 & \cellcolor[rgb]{ .886,  .627,  .463} 38.84 & \cellcolor[rgb]{ .918,  .706,  .573} 34.94 & \cellcolor[rgb]{ .776,  .349,  .067} \textbf{52.97} \\
          & 95.00 & \cellcolor[rgb]{ .988,  .894,  .839} 37.41 & \cellcolor[rgb]{ .941,  .773,  .667} 40.95 & \cellcolor[rgb]{ .902,  .675,  .525} 43.83 & \cellcolor[rgb]{ .776,  .349,  .067} \textbf{53.17} & \cellcolor[rgb]{ .988,  .894,  .839} 86.45 & \cellcolor[rgb]{ .894,  .647,  .486} 88.43 & \cellcolor[rgb]{ .902,  .667,  .514} 88.27 & \cellcolor[rgb]{ .776,  .349,  .067} \textbf{90.76} & \cellcolor[rgb]{ .988,  .894,  .839} 51.17 & \cellcolor[rgb]{ .933,  .745,  .627} 55.33 & \cellcolor[rgb]{ .898,  .659,  .506} 57.69 & \cellcolor[rgb]{ .776,  .349,  .067} \textbf{66.18} \\
\cmidrule{1-1}
    \multirow{5}[1]{*}{\begin{tabular}{c}\textbf{C3}\\ \textbf{average}\end{tabular}} & 99.95 & \cellcolor[rgb]{ .988,  .894,  .839} 1.04 & \cellcolor[rgb]{ .976,  .859,  .788} 1.46 & \cellcolor[rgb]{ .976,  .859,  .788} 1.44 & \cellcolor[rgb]{ .776,  .349,  .067} \textbf{6.96} & \cellcolor[rgb]{ .906,  .678,  .529} 94.80 & \cellcolor[rgb]{ .941,  .773,  .667} 93.56 & \cellcolor[rgb]{ .988,  .894,  .839} 91.97 & \cellcolor[rgb]{ .776,  .349,  .067} \textbf{99.01} & \cellcolor[rgb]{ .988,  .894,  .839} 2.05 & \cellcolor[rgb]{ .973,  .855,  .784} 2.85 & \cellcolor[rgb]{ .976,  .859,  .788} 2.78 & \cellcolor[rgb]{ .776,  .349,  .067} \textbf{12.85} \\
          & 99.90 & \cellcolor[rgb]{ .988,  .894,  .839} 1.20 & \cellcolor[rgb]{ .973,  .851,  .776} 1.78 & \cellcolor[rgb]{ .976,  .859,  .792} 1.65 & \cellcolor[rgb]{ .776,  .349,  .067} \textbf{8.04} & \cellcolor[rgb]{ .894,  .651,  .494} 91.31 & \cellcolor[rgb]{ .922,  .718,  .588} 89.73 & \cellcolor[rgb]{ .988,  .894,  .839} 85.41 & \cellcolor[rgb]{ .776,  .349,  .067} \textbf{98.50} & \cellcolor[rgb]{ .988,  .894,  .839} 2.37 & \cellcolor[rgb]{ .973,  .847,  .773} 3.46 & \cellcolor[rgb]{ .976,  .859,  .792} 3.18 & \cellcolor[rgb]{ .776,  .349,  .067} \textbf{14.76} \\
          & 99.50 & \cellcolor[rgb]{ .988,  .894,  .839} 4.12 & \cellcolor[rgb]{ .98,  .875,  .812} 4.74 & \cellcolor[rgb]{ .98,  .867,  .8} 4.92 & \cellcolor[rgb]{ .776,  .349,  .067} \textbf{19.50} & \cellcolor[rgb]{ .98,  .867,  .8} 85.62 & \cellcolor[rgb]{ .918,  .706,  .573} 89.05 & \cellcolor[rgb]{ .988,  .894,  .839} 85.01 & \cellcolor[rgb]{ .776,  .349,  .067} \textbf{96.63} & \cellcolor[rgb]{ .988,  .894,  .839} 7.83 & \cellcolor[rgb]{ .98,  .871,  .804} 8.96 & \cellcolor[rgb]{ .98,  .867,  .804} 9.03 & \cellcolor[rgb]{ .776,  .349,  .067} \textbf{31.65} \\
          & 99.00 & \cellcolor[rgb]{ .984,  .882,  .82} 7.40 & \cellcolor[rgb]{ .957,  .804,  .714} 9.89 & \cellcolor[rgb]{ .988,  .894,  .839} 6.92 & \cellcolor[rgb]{ .776,  .349,  .067} \textbf{24.81} & \cellcolor[rgb]{ .976,  .859,  .788} 83.64 & \cellcolor[rgb]{ .91,  .69,  .549} 87.41 & \cellcolor[rgb]{ .988,  .894,  .839} 82.80 & \cellcolor[rgb]{ .776,  .349,  .067} \textbf{94.99} & \cellcolor[rgb]{ .98,  .875,  .812} 13.51 & \cellcolor[rgb]{ .949,  .792,  .698} 17.32 & \cellcolor[rgb]{ .988,  .894,  .839} 12.48 & \cellcolor[rgb]{ .776,  .349,  .067} \textbf{38.32} \\
          & 95.00 & \cellcolor[rgb]{ .98,  .875,  .808} 24.82 & \cellcolor[rgb]{ .949,  .784,  .682} 27.35 & \cellcolor[rgb]{ .988,  .894,  .839} 24.18 & \cellcolor[rgb]{ .776,  .349,  .067} \textbf{39.79} & \cellcolor[rgb]{ .988,  .894,  .839} 80.99 & \cellcolor[rgb]{ .945,  .78,  .675} 82.36 & \cellcolor[rgb]{ .984,  .886,  .824} 81.13 & \cellcolor[rgb]{ .776,  .349,  .067} \textbf{87.38} & \cellcolor[rgb]{ .98,  .871,  .804} 37.59 & \cellcolor[rgb]{ .945,  .784,  .682} 40.28 & \cellcolor[rgb]{ .988,  .894,  .839} 36.73 & \cellcolor[rgb]{ .776,  .349,  .067} \textbf{53.82} \\
    \midrule
    \multirow{5}[2]{*}{\begin{tabular}{c}\textbf{Overall}\\ \textbf{AVERAGE}\end{tabular}} & 99.95 & \cellcolor[rgb]{ .988,  .894,  .839} 4.22 & \cellcolor[rgb]{ .937,  .765,  .655} 6.57 & \cellcolor[rgb]{ .918,  .71,  .576} 7.53 & \cellcolor[rgb]{ .776,  .349,  .067} \textbf{13.92} & \cellcolor[rgb]{ .867,  .576,  .388} 96.55 & \cellcolor[rgb]{ .843,  .518,  .306} 97.27 & \cellcolor[rgb]{ .988,  .894,  .839} 92.65 & \cellcolor[rgb]{ .776,  .349,  .067} \textbf{99.33} & \cellcolor[rgb]{ .988,  .894,  .839} 7.63 & \cellcolor[rgb]{ .933,  .749,  .631} 11.77 & \cellcolor[rgb]{ .922,  .718,  .588} 12.62 & \cellcolor[rgb]{ .776,  .349,  .067} \textbf{22.98} \\
          & 99.90 & \cellcolor[rgb]{ .988,  .894,  .839} 4.54 & \cellcolor[rgb]{ .937,  .757,  .643} 7.42 & \cellcolor[rgb]{ .914,  .702,  .565} 8.56 & \cellcolor[rgb]{ .776,  .349,  .067} \textbf{15.79} & \cellcolor[rgb]{ .906,  .675,  .525} 94.11 & \cellcolor[rgb]{ .863,  .565,  .373} 95.77 & \cellcolor[rgb]{ .988,  .894,  .839} 90.73 & \cellcolor[rgb]{ .776,  .349,  .067} \textbf{99.04} & \cellcolor[rgb]{ .988,  .894,  .839} 8.22 & \cellcolor[rgb]{ .929,  .741,  .62} 13.30 & \cellcolor[rgb]{ .918,  .71,  .576} 14.33 & \cellcolor[rgb]{ .776,  .349,  .067} \textbf{25.96} \\
          & 99.50 & \cellcolor[rgb]{ .988,  .894,  .839} 11.04 & \cellcolor[rgb]{ .937,  .757,  .643} 15.23 & \cellcolor[rgb]{ .91,  .69,  .549} 17.14 & \cellcolor[rgb]{ .776,  .349,  .067} \textbf{27.24} & \cellcolor[rgb]{ .988,  .894,  .839} 92.05 & \cellcolor[rgb]{ .898,  .655,  .498} 94.56 & \cellcolor[rgb]{ .976,  .859,  .788} 92.44 & \cellcolor[rgb]{ .776,  .349,  .067} \textbf{97.71} & \cellcolor[rgb]{ .988,  .894,  .839} 18.87 & \cellcolor[rgb]{ .929,  .745,  .627} 25.18 & \cellcolor[rgb]{ .914,  .702,  .565} 26.94 & \cellcolor[rgb]{ .776,  .349,  .067} \textbf{41.54} \\
          & 99.00 & \cellcolor[rgb]{ .988,  .894,  .839} 15.97 & \cellcolor[rgb]{ .929,  .741,  .62} 21.19 & \cellcolor[rgb]{ .91,  .694,  .553} 22.73 & \cellcolor[rgb]{ .776,  .349,  .067} \textbf{34.18} & \cellcolor[rgb]{ .988,  .894,  .839} 90.35 & \cellcolor[rgb]{ .894,  .647,  .49} 93.15 & \cellcolor[rgb]{ .961,  .82,  .733} 91.20 & \cellcolor[rgb]{ .776,  .349,  .067} \textbf{96.52} & \cellcolor[rgb]{ .988,  .894,  .839} 25.97 & \cellcolor[rgb]{ .925,  .725,  .6} 33.28 & \cellcolor[rgb]{ .918,  .706,  .573} 34.04 & \cellcolor[rgb]{ .776,  .349,  .067} \textbf{49.29} \\
          & 95.00 & \cellcolor[rgb]{ .984,  .878,  .816} 39.19 & \cellcolor[rgb]{ .988,  .894,  .839} 38.79 & \cellcolor[rgb]{ .894,  .651,  .494} 44.56 & \cellcolor[rgb]{ .776,  .349,  .067} \textbf{51.66} & \cellcolor[rgb]{ .988,  .894,  .839} 86.30 & \cellcolor[rgb]{ .953,  .8,  .702} 86.99 & \cellcolor[rgb]{ .933,  .745,  .627} 87.37 & \cellcolor[rgb]{ .776,  .349,  .067} \textbf{90.18} & \cellcolor[rgb]{ .988,  .894,  .839} 52.38 & \cellcolor[rgb]{ .988,  .886,  .827} 52.57 & \cellcolor[rgb]{ .91,  .694,  .553} 56.94 & \cellcolor[rgb]{ .776,  .349,  .067} \textbf{64.63} \\
    \bottomrule
    \end{tabular}%
\end{table}%

\begin{table}[h]
\footnotesize
  \centering
\caption{Area under curves. AUROC and AUPR curves are given for seven datasets and three types, C1, C2, C3, for each, as well as averages for each type and overall average across types. Darker colours represent better results. The best results are in bold. \label{table_AUC}} 
    \begin{tabular}{|c|cccc|cccc|}
    \toprule
    \multirow{3}[6]{*}{\textbf{Dataset}} & \multicolumn{4}{c|}{\textbf{AUROC}} & \multicolumn{4}{c|}{\textbf{AUPR}} \\
\cmidrule{2-9}          & \textbf{$\!\!\!$Martin$\!\!\!$} & \textbf{$\!\!\!$PIPE2$\!\!\!$} & \textbf{$\!\!\!$Ding$\!\!\!$} & \textbf{$\!\!\!$SPRINT$\!\!\!$} & \textbf{$\!\!\!$Martin$\!\!\!$} & \textbf{$\!\!\!$PIPE2$\!\!\!$} & \textbf{$\!\!\!$Ding$\!\!\!$} & \textbf{$\!\!\!$SPRINT$\!\!\!$} \\
\cmidrule{2-9}          & \multicolumn{8}{c|}{\textbf{C1}} \\
    \midrule
    \textbf{Biogrid} & \cellcolor[rgb]{ .863,  .565,  .373} 87.54 & \cellcolor[rgb]{ .988,  .894,  .839} 79.01 & \cellcolor[rgb]{ .776,  .349,  .067} \textbf{93.06} & \cellcolor[rgb]{ .855,  .541,  .341} 88.11 & \cellcolor[rgb]{ .878,  .608,  .431} 87.20 & \cellcolor[rgb]{ .988,  .894,  .839} 80.52 & \cellcolor[rgb]{ .776,  .349,  .067} \textbf{93.08} & \cellcolor[rgb]{ .843,  .518,  .306} 89.24 \\
    \textbf{HPRD} & \cellcolor[rgb]{ .847,  .525,  .318} 86.83 & \cellcolor[rgb]{ .988,  .894,  .839} 81.53 & \cellcolor[rgb]{ .776,  .349,  .067} \textbf{89.34} & \cellcolor[rgb]{ .847,  .529,  .325} 86.76 & \cellcolor[rgb]{ .898,  .655,  .498} 86.93 & \cellcolor[rgb]{ .988,  .894,  .839} 84.31 & \cellcolor[rgb]{ .776,  .349,  .067} \textbf{90.20} & \cellcolor[rgb]{ .812,  .431,  .184} 89.32 \\
    \textbf{Innate\_Exp} & \cellcolor[rgb]{ .855,  .553,  .353} 90.18 & \cellcolor[rgb]{ .988,  .894,  .839} 83.98 & \cellcolor[rgb]{ .776,  .349,  .067} \textbf{93.83} & \cellcolor[rgb]{ .831,  .49,  .263} 91.34 & \cellcolor[rgb]{ .871,  .592,  .412} 90.31 & \cellcolor[rgb]{ .988,  .894,  .839} 85.48 & \cellcolor[rgb]{ .776,  .349,  .067} \textbf{94.14} & \cellcolor[rgb]{ .824,  .471,  .239} 92.25 \\
    \textbf{Innate\_Man} & \cellcolor[rgb]{ .816,  .443,  .2} 94.11 & \cellcolor[rgb]{ .988,  .894,  .839} 90.26 & \cellcolor[rgb]{ .776,  .349,  .067} \textbf{94.89} & \cellcolor[rgb]{ .863,  .565,  .369} 93.09 & \cellcolor[rgb]{ .827,  .475,  .247} 94.93 & \cellcolor[rgb]{ .988,  .894,  .839} 92.22 & \cellcolor[rgb]{ .776,  .349,  .067} \textbf{95.73} & \cellcolor[rgb]{ .839,  .502,  .286} 94.75 \\
    \textbf{IntAct} & \cellcolor[rgb]{ .855,  .549,  .349} 88.02 & \cellcolor[rgb]{ .988,  .894,  .839} 80.72 & \cellcolor[rgb]{ .776,  .349,  .067} \textbf{92.18} & \cellcolor[rgb]{ .843,  .518,  .302} 88.69 & \cellcolor[rgb]{ .875,  .596,  .42} 87.51 & \cellcolor[rgb]{ .988,  .894,  .839} 81.68 & \cellcolor[rgb]{ .776,  .349,  .067} \textbf{92.31} & \cellcolor[rgb]{ .831,  .486,  .259} 89.71 \\
    \textbf{MINT} & \cellcolor[rgb]{ .835,  .494,  .275} 90.86 & \cellcolor[rgb]{ .988,  .894,  .839} 83.41 & \cellcolor[rgb]{ .776,  .349,  .067} \textbf{93.54} & \cellcolor[rgb]{ .875,  .592,  .412} 89.03 & \cellcolor[rgb]{ .859,  .553,  .357} 91.08 & \cellcolor[rgb]{ .988,  .894,  .839} 85.93 & \cellcolor[rgb]{ .776,  .349,  .067} \textbf{94.11} & \cellcolor[rgb]{ .855,  .549,  .349} 91.13 \\
    \textbf{Park \& Marcotte} & \cellcolor[rgb]{ .812,  .435,  .188} 81.49 & \cellcolor[rgb]{ .988,  .894,  .839} 76.74 & \cellcolor[rgb]{ .792,  .384,  .118} 82.00 & \cellcolor[rgb]{ .776,  .349,  .067} \textbf{82.35} & \cellcolor[rgb]{ .898,  .655,  .502} 82.32 & \cellcolor[rgb]{ .988,  .894,  .839} 79.90 & \cellcolor[rgb]{ .871,  .588,  .404} 83.00 & \cellcolor[rgb]{ .776,  .349,  .067} \textbf{85.39} \\
    \midrule
          & \multicolumn{8}{c|}{\textbf{C2}} \\
    \midrule
    \textbf{Biogrid} & \cellcolor[rgb]{ .89,  .639,  .478} 81.33 & \cellcolor[rgb]{ .988,  .894,  .839} 76.66 & \cellcolor[rgb]{ .776,  .349,  .067} \textbf{86.57} & \cellcolor[rgb]{ .82,  .455,  .216} 84.67 & \cellcolor[rgb]{ .925,  .725,  .6} 80.76 & \cellcolor[rgb]{ .988,  .894,  .839} 78.25 & \cellcolor[rgb]{ .784,  .365,  .086} 86.12 & \cellcolor[rgb]{ .776,  .349,  .067} \textbf{86.30} \\
    \textbf{HPRD} & \cellcolor[rgb]{ .91,  .686,  .545} 83.30 & \cellcolor[rgb]{ .988,  .894,  .839} 81.55 & \cellcolor[rgb]{ .839,  .506,  .29} 84.78 & \cellcolor[rgb]{ .776,  .349,  .067} \textbf{86.09} & \cellcolor[rgb]{ .988,  .894,  .839} 82.85 & \cellcolor[rgb]{ .945,  .784,  .682} 83.98 & \cellcolor[rgb]{ .914,  .698,  .561} 84.85 & \cellcolor[rgb]{ .776,  .349,  .067} \textbf{88.37} \\
    \textbf{Innate\_Exp} & \cellcolor[rgb]{ .922,  .722,  .596} 83.96 & \cellcolor[rgb]{ .988,  .894,  .839} 81.46 & \cellcolor[rgb]{ .816,  .443,  .2} 87.98 & \cellcolor[rgb]{ .776,  .349,  .067} \textbf{89.31} & \cellcolor[rgb]{ .957,  .816,  .725} 83.74 & \cellcolor[rgb]{ .988,  .894,  .839} 82.57 & \cellcolor[rgb]{ .847,  .522,  .314} 87.91 & \cellcolor[rgb]{ .776,  .349,  .067} \textbf{90.37} \\
    \textbf{Innate\_Man} & \cellcolor[rgb]{ .894,  .651,  .494} 85.87 & \cellcolor[rgb]{ .988,  .894,  .839} 84.43 & \cellcolor[rgb]{ .969,  .843,  .769} 84.74 & \cellcolor[rgb]{ .776,  .349,  .067} \textbf{87.64} & \cellcolor[rgb]{ .949,  .792,  .694} 87.71 & \cellcolor[rgb]{ .98,  .875,  .812} 87.22 & \cellcolor[rgb]{ .988,  .894,  .839} 87.10 & \cellcolor[rgb]{ .776,  .349,  .067} \textbf{90.33} \\
    \textbf{IntAct} & \cellcolor[rgb]{ .882,  .62,  .451} 81.68 & \cellcolor[rgb]{ .988,  .894,  .839} 77.64 & \cellcolor[rgb]{ .776,  .349,  .067} \textbf{85.63} & \cellcolor[rgb]{ .843,  .522,  .31} 83.14 & \cellcolor[rgb]{ .929,  .737,  .62} 80.68 & \cellcolor[rgb]{ .988,  .894,  .839} 78.69 & \cellcolor[rgb]{ .788,  .38,  .11} 85.20 & \cellcolor[rgb]{ .776,  .349,  .067} \textbf{85.58} \\
    \textbf{MINT} & \cellcolor[rgb]{ .8,  .4,  .141} 86.66 & \cellcolor[rgb]{ .988,  .894,  .839} 81.76 & \cellcolor[rgb]{ .776,  .349,  .067} \textbf{87.17} & \cellcolor[rgb]{ .816,  .447,  .208} 86.20 & \cellcolor[rgb]{ .878,  .612,  .439} 86.37 & \cellcolor[rgb]{ .988,  .894,  .839} 84.08 & \cellcolor[rgb]{ .827,  .475,  .243} 87.47 & \cellcolor[rgb]{ .776,  .349,  .067} \textbf{88.47} \\
    \textbf{Park \& Marcotte} & \cellcolor[rgb]{ .965,  .831,  .749} 60.67 & \cellcolor[rgb]{ .847,  .525,  .314} 63.76 & \cellcolor[rgb]{ .988,  .894,  .839} 60.00 & \cellcolor[rgb]{ .776,  .349,  .067} \textbf{65.52} & \cellcolor[rgb]{ .98,  .875,  .808} 60.43 & \cellcolor[rgb]{ .835,  .502,  .282} 67.41 & \cellcolor[rgb]{ .988,  .894,  .839} 60.00 & \cellcolor[rgb]{ .776,  .349,  .067} \textbf{70.25} \\
    \midrule
          & \multicolumn{8}{c|}{\textbf{C3}} \\
    \midrule
    \textbf{Biogrid} & \cellcolor[rgb]{ .867,  .58,  .392} 76.20 & \cellcolor[rgb]{ .988,  .894,  .839} 71.38 & \cellcolor[rgb]{ .792,  .384,  .118} 79.16 & \cellcolor[rgb]{ .776,  .349,  .067} \textbf{79.67} & \cellcolor[rgb]{ .894,  .651,  .494} 74.89 & \cellcolor[rgb]{ .988,  .894,  .839} 70.25 & \cellcolor[rgb]{ .847,  .525,  .318} 77.24 & \cellcolor[rgb]{ .776,  .349,  .067} \textbf{80.59} \\
    \textbf{HPRD} & \cellcolor[rgb]{ .91,  .69,  .549} 79.46 & \cellcolor[rgb]{ .988,  .894,  .839} 77.14 & \cellcolor[rgb]{ .976,  .863,  .792} 77.51 & \cellcolor[rgb]{ .776,  .349,  .067} \textbf{83.27} & \cellcolor[rgb]{ .922,  .718,  .588} 78.51 & \cellcolor[rgb]{ .925,  .729,  .608} 78.28 & \cellcolor[rgb]{ .988,  .894,  .839} 75.32 & \cellcolor[rgb]{ .776,  .349,  .067} \textbf{85.08} \\
    \textbf{Innate\_Exp} & \cellcolor[rgb]{ .941,  .773,  .667} 78.10 & \cellcolor[rgb]{ .988,  .894,  .839} 75.89 & \cellcolor[rgb]{ .886,  .627,  .463} 80.69 & \cellcolor[rgb]{ .776,  .349,  .067} \textbf{85.70} & \cellcolor[rgb]{ .949,  .792,  .694} 76.65 & \cellcolor[rgb]{ .988,  .894,  .839} 74.42 & \cellcolor[rgb]{ .918,  .706,  .573} 78.55 & \cellcolor[rgb]{ .776,  .349,  .067} \textbf{86.23} \\
    \textbf{Innate\_Man} & \cellcolor[rgb]{ .875,  .6,  .42} 71.75 & \cellcolor[rgb]{ .843,  .522,  .31} 73.25 & \cellcolor[rgb]{ .988,  .894,  .839} 65.96 & \cellcolor[rgb]{ .776,  .349,  .067} \textbf{76.57} & \cellcolor[rgb]{ .886,  .624,  .455} 73.49 & \cellcolor[rgb]{ .863,  .565,  .369} 74.95 & \cellcolor[rgb]{ .988,  .894,  .839} 66.81 & \cellcolor[rgb]{ .776,  .349,  .067} \textbf{80.17} \\
    \textbf{IntAct} & \cellcolor[rgb]{ .855,  .549,  .345} 76.94 & \cellcolor[rgb]{ .988,  .894,  .839} 73.61 & \cellcolor[rgb]{ .776,  .349,  .067} \textbf{78.81} & \cellcolor[rgb]{ .957,  .808,  .718} 74.44 & \cellcolor[rgb]{ .914,  .702,  .565} 74.88 & \cellcolor[rgb]{ .988,  .894,  .839} 73.11 & \cellcolor[rgb]{ .867,  .576,  .388} 76.03 & \cellcolor[rgb]{ .776,  .349,  .067} \textbf{78.08} \\
    \textbf{MINT} & \cellcolor[rgb]{ .839,  .506,  .29} 81.25 & \cellcolor[rgb]{ .988,  .894,  .839} 78.06 & \cellcolor[rgb]{ .949,  .788,  .69} 78.94 & \cellcolor[rgb]{ .776,  .349,  .067} \textbf{82.54} & \cellcolor[rgb]{ .906,  .678,  .533} 80.07 & \cellcolor[rgb]{ .929,  .737,  .62} 79.28 & \cellcolor[rgb]{ .988,  .894,  .839} 77.14 & \cellcolor[rgb]{ .776,  .349,  .067} \textbf{84.55} \\
    \textbf{Park \& Marcotte} & \cellcolor[rgb]{ .941,  .765,  .655} 57.86 & \cellcolor[rgb]{ .878,  .608,  .435} 58.90 & \cellcolor[rgb]{ .988,  .894,  .839} 57.00 & \cellcolor[rgb]{ .776,  .349,  .067} \textbf{60.60} & \cellcolor[rgb]{ .961,  .82,  .729} 57.07 & \cellcolor[rgb]{ .882,  .616,  .447} 59.84 & \cellcolor[rgb]{ .988,  .894,  .839} 56.00 & \cellcolor[rgb]{ .776,  .349,  .067} \textbf{63.49} \\
    \midrule
          & \multicolumn{8}{c|}{\textbf{AVERAGES}} \\
    \midrule
    \textbf{C1 average} & \cellcolor[rgb]{ .843,  .522,  .31} 88.43 & \cellcolor[rgb]{ .988,  .894,  .839} 82.24 & \cellcolor[rgb]{ .776,  .349,  .067} \textbf{91.26} & \cellcolor[rgb]{ .843,  .518,  .306} 88.48 & \cellcolor[rgb]{ .867,  .584,  .396} 88.61 & \cellcolor[rgb]{ .988,  .894,  .839} 84.29 & \cellcolor[rgb]{ .776,  .349,  .067} \textbf{91.80} & \cellcolor[rgb]{ .824,  .463,  .227} 90.26 \\
\cmidrule{1-1}    \textbf{C2 average} & \cellcolor[rgb]{ .894,  .647,  .486} 80.50 & \cellcolor[rgb]{ .988,  .894,  .839} 78.18 & \cellcolor[rgb]{ .812,  .439,  .192} 82.41 & \cellcolor[rgb]{ .776,  .349,  .067} \textbf{83.23} & \cellcolor[rgb]{ .988,  .89,  .835} 80.36 & \cellcolor[rgb]{ .988,  .894,  .839} 80.32 & \cellcolor[rgb]{ .898,  .655,  .502} 82.67 & \cellcolor[rgb]{ .776,  .349,  .067} \textbf{85.67} \\
\cmidrule{1-1}    \textbf{C3 average} & \cellcolor[rgb]{ .91,  .686,  .545} 74.51 & \cellcolor[rgb]{ .988,  .894,  .839} 72.60 & \cellcolor[rgb]{ .929,  .741,  .62} 74.01 & \cellcolor[rgb]{ .776,  .349,  .067} \textbf{77.54} & \cellcolor[rgb]{ .957,  .804,  .714} 73.65 & \cellcolor[rgb]{ .976,  .863,  .796} 72.88 & \cellcolor[rgb]{ .988,  .894,  .839} 72.44 & \cellcolor[rgb]{ .776,  .349,  .067} \textbf{79.74} \\
    \midrule
    \textbf{Overall AVERAGE} & \cellcolor[rgb]{ .855,  .545,  .345} 81.15 & \cellcolor[rgb]{ .988,  .894,  .839} 77.67 & \cellcolor[rgb]{ .8,  .404,  .141} 82.56 & \cellcolor[rgb]{ .776,  .349,  .067} \textbf{83.08} & \cellcolor[rgb]{ .929,  .741,  .624} 80.87 & \cellcolor[rgb]{ .988,  .894,  .839} 79.16 & \cellcolor[rgb]{ .882,  .612,  .439} 82.30 & \cellcolor[rgb]{ .776,  .349,  .067} \textbf{85.22} \\
    \bottomrule
    \end{tabular}%
\end{table}%

\subsection{Predicting the entire human interactome}

The goal of all PPI prediction methods is to predict new interactions from existing reliable ones. That means, in practice we input all known interactions -- the entire interactome of an organism -- and predict new ones. Of the newly predicted intreactions, only those that are the most likely to be true interactions are kept. 

For predicting the entire interactome, we need to predict the probability of interaction between any two proteins. For $N$ proteins, that means we need to consider $(N^2+N)/2$ protein pairs. For our 20,160 proteins, that is about 203 million potential interactions. 
For example, predicting one pair per second results in over six years of computation time. 

We have tested the four programs, Martin's, PIPE2, Ding's, and SPRINT, on the entire human interactome, considering as given PPIs each of the six datasets in Table~\ref{table_datasets}. The tests were performed on a DELL PowerEdge R620 computer with 12 cores Intel Xeon at 2.0 GHz and 256 GB of RAM, running Linux Red Hat, CentOS 6.3. 

The time and memory values are shown in Table~\ref{table_interactome_time} for all three stages: preprocessing, training, and predicting. For each dataset, training is performed on all PPIs in that dataset and then predictions are made for all 203 million protein pairs. 

Note that PIPE2 and SPRINT do not require any training. Also, preprocessing is performed only once for all protein sequences. As long as no protein sequences are added, no preprocessing needs to be done. For SPRINT, we provide all necessary similarities for all reviewed human proteins in UniProt. If new protein sequences are added, the code can be easily modified to preprocess only those.

Therefore, the comparison is between predicting time of PIPE2 and SPRINT and total time of Martin and Ding. PIPE2 and Martin are very slow and the predicting times are estimated by running the programs for 100 hours and then estimating according to the number of protein pairs left to process. Both take too long to be used on the entire human interactome.

Ding's program is much faster but uses a large amount of memory. It ran out of 256GB of memory when training on the two largest datasets: Biogrid and InnateDB experimentally validated. It seems able to train on the IntAct dataset but it could not finish training in 14 days, which is the longest we can run a job on our system.

SPRINT is approximately five orders of magnitude faster than PIPE2 and Martin. It is over two orders of magnitude faster than Ding but this is based on the small datasets. The results on IntAct seem to indicate that the difference is larger for large datasets. 

It should be noted that SPRINT runs in parallel whereas the other are serial. However, the difference is large anyway. 

In terms of memory, SPRINT requires a very modest amount of memory to predict. We successfully ran SPRINT on all entire human interactome tests in serial mode on an older MacBook (1.4GHz processor, 4 GB RAM); the running time was between 35 minutes for Innate manually curated to 11 hours for Biogriod.

The very large difference in speed is due to the fact that while Martin, PIPE2, and Ding consider one protein pair at the time, out of the 203 million, SPRINT simply computes all 203 million scores at the same time; see the Methods section for details.

\begin{table}[h]
\centering
\caption{\textbf{Human interactome comparison: running time and peak memory.} The predicting time for Martin's and PIPE2 was estimated by running it for 100 hours and then estimating the total time according to the number of pairs left to predict. Note that PIPE2 and SPRINT do not require training as they are not using machine learning. For the entries marked with a dash, the program ran out of (256 GB) memory or ran for more than 14 days. Times marked with a dagger${}^\dag$ are estimated.\label{table_interactome_time}}
\begin{tabular}{@{}llrrrrrr@{}} \toprule
Dataset & Program & \multicolumn{3}{c}{Time (s)} &  \multicolumn{3}{|c}{Memory (GB)}   \\ \cmidrule{3-5} \cmidrule{6-8}
 &	& Preprocess & Train & Predict &  Preprocess & Train & Predict \\ \midrule
Biogrid & 
	Martin & 32,400 & $>$ 1,209,600  & {\scriptsize  -- } & 2.5 & 6.1 & {\scriptsize  -- } \\ 
&	PIPE2 & 312,120 & {\scriptsize N$\!$/$\!$A} & ${}^\dag$1,150,675,200 & 2.1 & {\scriptsize N$\!$/$\!$A} & 18.9\\ 
&	Ding	   & 37,708 & {\scriptsize  -- } & {\scriptsize  -- }  & 3.3 &  $>$ 256 & {\scriptsize  -- }\\
&	SPRINT & 33,480 & {\scriptsize N$\!$/$\!$A} & 6,120 & 11.2 & {\scriptsize N$\!$/$\!$A} & 3.0 \\ \midrule
HPRD  & 
	Martin & 32,400 & 584,640 & ${}^\dag$107,222,400  & 2.5 & 3.2 & 1.5 \\ 
Release 9 &	PIPE2 & 312,120 & {\scriptsize N$\!$/$\!$A} & ${}^\dag$435,628,800 &  2.1 & {\scriptsize N$\!$/$\!$A} & 18.9 \\ 
&	Ding	   & 37,708 & 236,551 & 374,360 & 3.3 & 79.5 & 79.5\\
&	SPRINT & 33,480 & {\scriptsize N$\!$/$\!$A} & 1,257 & 11.2 & {\scriptsize N$\!$/$\!$A} & 3.0 \\ \midrule
Innate   & 
	Martin & 32,400 & $>$ 1,209,600 & {\scriptsize  -- }  & 2.5 & 5.7 & {\scriptsize  -- } \\ 
experim. &	PIPE2 & 312,120 & {\scriptsize N$\!$/$\!$A} & ${}^\dag$872,294,400 &  2.1 &  {\scriptsize N$\!$/$\!$A} & 18.9 \\ 
validated &	Ding	   & 37,708 & {\scriptsize  -- } & {\scriptsize  -- }  & 3.3 &  $>$ 256 & {\scriptsize  -- } \\
&	SPRINT & 33,480 & {\scriptsize N$\!$/$\!$A} & 3,600 & 11.2 & {\scriptsize N$\!$/$\!$A} & 3.0  \\ \midrule
Innate  &  
	Martin & 32,400 & 26,280 & ${}^\dag$30,888,000 & 2.5 & 1.9 & 1.5 \\ 
manually &	PIPE2 & 312,120 & {\scriptsize N$\!$/$\!$A} & ${}^\dag$230,342,400  &  2.1 &  {\scriptsize N$\!$/$\!$A} & 18.9 \\ 
curated &	Ding	   & 37,708 & 55,532 & 285,323 & 3.3 & 25.4 & 25.4 \\
&	SPRINT & 33,480 & {\scriptsize N$\!$/$\!$A} & 930 & 11.2 & {\scriptsize N$\!$/$\!$A} & 3.0  \\ \midrule
IntAct & 
	Martin & 32,400 & $>$ 1,209,600 & {\scriptsize  -- }    & 2.5 & 3.5 & {\scriptsize  -- } \\ 
&	PIPE2 & 312,120 & {\scriptsize N$\!$/$\!$A} & ${}^\dag$616,464,000 &  2.1 &  {\scriptsize N$\!$/$\!$A} & 18.9\\ 
&	Ding	   & 37,708 &  $>$ 1,209,600 & {\scriptsize  -- }   & 3.3 & 220 & {\scriptsize  -- } \\
&	SPRINT & 33,480 & {\scriptsize N$\!$/$\!$A} & 2,672 & 11.2 & {\scriptsize N$\!$/$\!$A} & 3.0  \\ \midrule
MINT & 
	Martin & 32,400 & 101,160  & ${}^\dag$52,557,120 & 2.5 & 2.3 & 1.5\\ 
&	PIPE2 & 312,120 & {\scriptsize N$\!$/$\!$A} & ${}^\dag$372,902,400 &  2.1 &  {\scriptsize N$\!$/$\!$A} & 18.9 \\ 
&	Ding	   & 37,708 & 120,720 & 331,865 & 3.3 & 41.1 & 41.1\\
&	SPRINT & 33,480 & {\scriptsize N$\!$/$\!$A} & 952 & 11.2 & {\scriptsize N$\!$/$\!$A} & 3.0  \\ \bottomrule
\end{tabular}
\end{table}


\section{Conclusion}

We have presented a new algorithm and software, SPRINT, for predicting PPIs that has higher performance than the current state-of-the-art programs while running orders of magnitude faster and using very little memory. SPRINT is very easy to use and we hope it will make PPI prediction for entire interactomes a routine task. It can be used on its own or in connection with other tools for PPI prediction.

Plenty of room for improvement remains, especially for the C2 and C3 data. Also, we hope to use the algorithm of SPRINT to predict interacting sites. Since they work directly with the sequence of amino acids, sequence-based methods often have an advantage in finding the actual positions where interaction occurs.


\section{Methods}

\subsection{Basic idea}

Proteins similar with interacting proteins are likely to interact as well. That is, if $P_1$ is known to interact with $P_2$ and the sequences of $P_1$ and $P'_1$ are highly similar and the sequences of $P_2$ and $P'_2$ are highly similar, then $P'_1$ and $P'_2$ are likely to interact as well. In a way or another, this is essentially the idea behind the brute force calculation of PIPE as well as the machine learning algorithms of Martin, Shen, and Guo. 

\begin{figure}
\centering
\includegraphics[width=8cm]{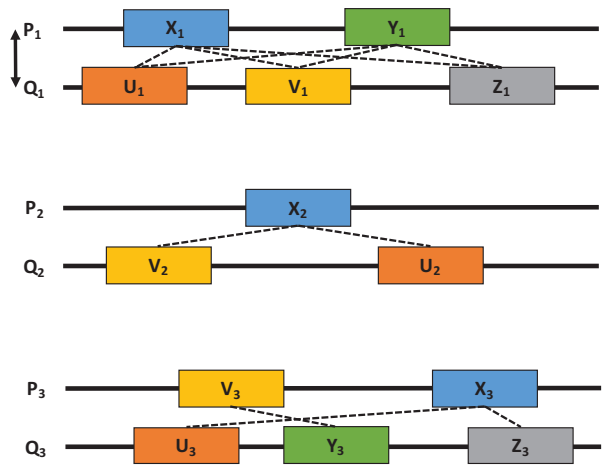}
\caption{\textbf{Interaction inference.} The proteins $P_1$ and $Q_1$ are known to interact; blocks of the same colour represent occurrences of similar subsequences. Dashed lines indicate potential contributions to interactions: there are six between $P_1$ and $Q_1$ and they imply two between $P_2$ and $Q_2$ and three between $P_3$ and $Q_3$. \label{fig_idea}}
\end{figure}

SPRINT uses a complex algorithm to quickly evaluate the contribution of similar subsequences to the likelihood of interaction. The basic idea is illustrated on a toy example in Figure~\ref{fig_idea}. Assume we have given three protein pairs $(P_1,Q_1)$, $(P_2,Q_2)$, $(P_3,Q_3)$, of which $(P_1,Q_1)$ is a known interaction. Also, assume that we have detected the similar subsequences indicated by blocks of the same colour in the figure. That is, $X_1, X_2$, and $X_3$ are similar with each other, $Y_1$ and $Y_3$ are similar, etc. In this context, the fact that $X_1$ and $U_1$ belong to interacting proteins increases the likelihood that $P_2$ and $Q_2$ interact because $P_2$ contains $X_2$ that is similar with $X_1$ and $Q_2$ contains $U_2$ that is similar with $U_1$. Six such subsequence pairs between the interacting proteins $P_1$ and $Q_1$ are marked with dashed lines in Figure~\ref{fig_idea} and they imply, using the above reasoning, two subsequence pairs in-between $P_2$ and $Q_2$ and three in between $P_3$ and $Q_3$, also marked with dashed lines. SPRINT is counting the contribution from such dash lines in order to estimate the likelihood of interaction of any protein pair. In our example, SPRINT would count two dash lines for $(P_2,Q_2)$ and three for $(P_3,Q_3)$.

Long similar regions should have a higher weight than short ones. To account for this we assume that all contributing blocks have a fixed length $k$ 
and that a region of length $\ell$ contributes $\ell-k+1$ blocks. As $k$ is fixed, this grows linearly with $\ell$. The precise score is  given later in this section.

\subsection{Finding similar subsequences}

As described above, the first step of SPRINT is the identification of similar subsequences among the input protein sequences. This is done using spaced seeds. Spaced seeds \cite{Ma02_PatternHunter,Li04_PatternHunterII} are an alternative to BLAST's hit-and-extend method, that we briefly recall. Assume a match of size five is used. In this case, an exact match consists of five consecutive matching amino acids between two protein sequences. This is called a {\it hit}. Any such hit is then {\it extended} to the left and to the right until the score drops below a given threshold. If the score is sufficiently high, then the two extended subsequences are reported as similar.

Denote the five consecutive matches of BLAST by {\tt 11111}; this is called a consecutive {\it seed} of weight five. {\it Spaced seeds} consists of matches interspersed by don't care positions; here is an example of such a spaced seed: {\tt 11****11***1}. A {\it spaced match} requires only the amino acids in positions corresponding to {\tt 1}'s in the seed to match; in the given example, only the amino acids in positions 1, 2, 7, 8, and 12 have to match. Note that the number of matches (the weight) is the same as for the consecutive seed; five in our case. There is a trade-off between speed and probability of finding similarities. Lower weight has increased sensitivity because it is easier to hit similar regions but lower speed since more random hits are expected and have to be processed. The best value for our problem turned out to be five.

\begin{figure}[h]
\centering
\includegraphics{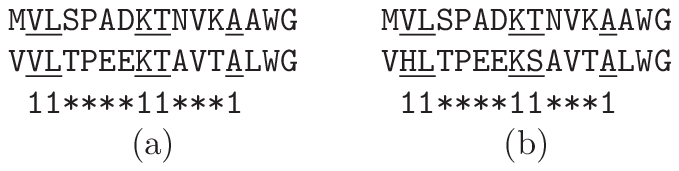}
\caption{An exact hit (a) and an approximate hit (b) of the same spaced seed.\label{fig_s-match}}
\end{figure}

The hit-and-extend approach works in the same way as described above, except that the initial matches are spaced as opposed to consecutive. An example of a hit is shown below. Given the spaced seed above, two \textit{exact} spaced matches are underlined in Figure~\ref{fig_s-match}(a).

Spaced seeds have higher probability of detecting similar subsequences, while the number of hits is the same as for consecutive seeds; the expected number of hits is given by the weight of the seed, which is the same; see \cite{Ma02_PatternHunter} for details. Several seeds \cite{Li04_PatternHunterII} can detect more similar subsequences as they capture different similarities. The distribution of matches and don't care positions is crucial for the quality of the seeds and we have used SpEED~\cite{Ilie07_spacedSeeds,Ilie11_SpEED} to compute the following seeds used by SPRINT; we have experimentally determined that four seeds of weight five are the best choice: $S_{4,5} = \{$\texttt{11****11***1}, \texttt{1**1*1***1*1}, \texttt{11**1***1**1}, \texttt{1*1******111}$\}$.

In order to further increase the probability of finding similar subsequences, we consider also hits between similar matches, as opposed to exact ones. For example, the following two amino acid sequences, though similar, do not have any \textit{exact} spaced matches. In order to capture such similarities, we consider also hits consisting of \textit{similar} spaced matches; an example is shown in Figure~\ref{fig_s-match}(b). All such hits of similar spaced matches are found using the spaced seeds shown above and then extended both ways in order to identify similar regions. Details of the fast implementation are given next.

\subsection{Implementation}

The protein sequences are encoded into bits using five bits per amino acid. (The five bits used for encoding are unrelated with the weight of the spaced seeds employed. It is a coincidence that both numbers are five. Each protein sequence is encoded as an array of unsigned 64-bit integers; each 64-bit integer stores 12 amino acids within 60 bits and 4 bits are unused. Each spaced seed is encoded using also five bits per position, {\tt 11111} for a {\tt 1} (match) and {\tt 00000} for a {\tt *} (don't care). Bitwise operations are then heavily used in order to speed up recording spaced-mers into hash tables. {\it Spaced-mers} are defined analogously with $k$-mers but using a spaced seed. A $k$-mer is a contiguous sequence of $k$ amino acids. Given a spaced seed, a spaced-mer consists of $k$ amino acids interspersed with spaces, according to the seed. For a spaced seed $s$, we shall call the spaced-mers also $s$-mers. Figure~\ref{fig_s-mers} shows an example of all $s$-mers of a sequence, for $s=${\tt 11****11***1}:

\begin{figure}[h]
\centering
\includegraphics{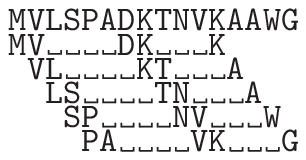}
\caption{An example of all $s$-mers of a sequence.\label{fig_s-mers}}
\end{figure}

All spaced-mers in all protein sequences are computed and stored in a hash table, together with their location in the protein sequences. Because of our representation, the computation of each spaced-mer requires only one bitwise AND and one bit SHIFT operation. Once all spaced-mers are stored, for each spaced-mer in the table, all similar spaced-mers are computed and then all hits between the spaced-mer and similar ones are easily collected from the table and extended in search for similarities.

\subsection{Post-processing similarities\label{sec_post_proc}}

\begin{figure}[h]
\centering
\includegraphics{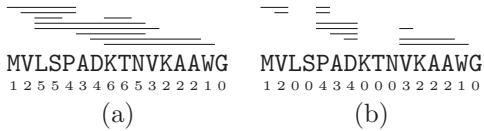}
\caption{An example of similarities before (a) and after (b) post-processing.\label{fig_similarities}}
\end{figure}

We first process the similar subsequences we computed in the previous phase to remove those appearing too many times as they are believed to be just repeats that occur very often in the protein sequences without any relevance for the interaction process. We explain the algorithm on the toy example below. For the protein sequence {\tt MVLSPADKTNVKAAWG}, assume we have found the similarities marked by lines in Figure~\ref{fig_similarities}(a). For example, the top line means that {\tt MVLSP} was found to be similar with another subsequence somewhere else, the bottom line represents the same about the subsequence {\tt KTNVKAAW}, etc.

The counts in the bottom row indicate how many times each position occurs in all similarities found. (In the figure above, this means the number of lines that cover that position.) All positions with a count above a threshold will be eliminated from all similarities, which will be modified accordingly. In our example, assuming the threshold is 5, positions 3, 4, 8, 9, and 10 have counts 5 or higher and are eliminated; see Figure~\ref{fig_similarities}(b). The new similarities are indicated by the lines above the sequence. For example, {\tt MVLSP} has positions 3 and 4 removed and becomes two similarities, {\tt MV} and {\tt P}. The counterpart of each similarity is modified the same way.

\subsection{Scoring PPIs}

What we have computed so far are similarities, that is, pairs of similar subsequences of the same length. We now show how to compute the scores. The score between two $k$-mers $A$ and $B$ is computed as the sum of all scores of corresponding amino acids ($S_e$ stands for the score between sequences of equal length):
\begin{equation}\label{eq_score_kmer}
S_e(A,B) = \sum_{i=1}^kM(A_i,B_i)\ ,
\end{equation}
where $A_i$ is the $i$th amino acid of $A$ and $M(A_i, B_i)$ is the score between amino acids $A_i$ and $B_i$ as given by the scoring matrix $M$. SPRINT can use any matrix $M$; by default it uses PAM120.

We now extend the definition of the score from $k$-mers to arbitrary subsequences.
For two subsequences $X$ and $Y$ of length $n$, the score is given by the sum of the scores of all corresponding $k$-mer pairs; using \eqref{eq_score_kmer}:
\begin{equation}\label{eq_score_nmer}
S_e(X,Y) = \sum_{i=1}^{n-k+1}S_e(X[i\pp i+k-1],Y[i\pp i+k-1])\ ,
\end{equation}
where $X[i\pp j] = X_iX_{i+1}\cdots X_j$. It is important to recall that any two similar sequences we find have the same length, therefore the above formula is correct.

Finally, we describe how the scores for whole protein sequences are computed. Initially all scores are set to zero. Each pair of proteins $(P_1, P_2)$ that are known to interact has its own contribution to the scores of other pairs. For each computed similarity $(X_1,Y_1)$ between $P_1$ and another protein $Q_1$ ($X_1$ is a subsequence of $P_1$ and $Y_1$ is a subsequence of $Q_1$) and for each similarity $(X_2,Y_2)$ between $P_2$ and another protein $Q_2$, the score between $Q_1$ and $Q_2$, $S(Q_1,Q_2)$, is increased, using \eqref{eq_score_nmer}, by:
\begin{equation} \label{eq_score_add}
\frac{S_e(X_1,Y_1)(|X_2|-k+1)+ S_e(X_2,Y_2)(|X_1|-k+1)}{|Q_1||Q_2|}\ ,
\end{equation}
where $|Q|$ denotes the length of the amino acid sequence $Q$. That means, the score of each corresponding $k$-mer pair between $X_1$ and $Y_1$ is multiplied by the number of $k$-mers in $X_2$, that is, the number of times it is used to support the fact that $Q_1$ is interacting with $Q_2$. Similarly, the score of each corresponding $k$-mer pair between $X_2$ and $Y_2$ is multiplied by the number of $k$-mers in $X_1$. The score obtained this way is then normalized by dividing it by the lengths of the proteins involved.

\subsection{Predicting interactions}

Once the score are computed, by considering all given interactions and similar subsequences and computing their impact on the other scores as above, predicting interactions is simply done according to the scores. All protein pairs are sorted decreasingly by the scores; higher score represent higher probability to interact. If a threshold is provided, then those pairs with scores above the threshold are reported as interacting. 

\subsection{SPRINT}

We put all the above together to summarize the SPRINT algorithm for predicting the entire interactome. 
The input consists of the proteins sequences and PPIs. The default set of seeds is given by $S_{4,5}$ above but any set can be used.

\smallskip

\noindent
\setcounter{algccc}{0}
\underline{$\text{\sc SPRINT}(P_s,P_i)$}\\
\textbf{input:} protein sequences $P_s$, protein interactions $P_i$\\
\textbf{global:} seed set $S$\\
\textbf{output:} all protein pairs sorted decreasingly by score\\
\   [Hash spaced-mers]\\
\ccc  \textbf{for} each seed $s$ in $S$ \textbf{do}\\
\ccc  \qqa \textbf{for} each protein sequence $p$ in $P_s$ \textbf{do}\\
\ccc  \qqb \textbf{for} $i$ \textbf{from} $0$ \textbf{to} $|p| - |s|$ \textbf{do}   \\
\ccc  \qqc $w$ $\gets$ the $s$-mer at position $i$ in $p$\\
\ccc  \qqc store $w$ in hash table $H_s$\\
\ccc  \qqc store $i$ in the list of $w$ [each $s$-mer has a list of positions]\\
\   [Compute similarities]\\
\ccc  \textbf{for} each seed $s$ in $S$ \textbf{do}\\
\ccc  \qqa \textbf{for} each $s$-mer $w$ in $H_s$ \textbf{do}\\
\ccc  \qqb compute the set $S_w$ of $s$-mers similar with $w$\\
\ccc  \qqb \textbf{for} each $z \in S_w$ \textbf{do}\\
\ccc  \qqc \textbf{for} each position $i$ in the list of $w$ \textbf{do}\\
\ccc  \qqd \textbf{for} each position $j$ in the list of $z$ \textbf{do}\\
\ccc  \qqe extend the similarity between $w$ and $z$ both ways   \\
\ccc  \qqe \textbf{if} similarity sufficiently high \\
\ccc  \qqf \textbf{then} store the pair of subsequences found \\
\ccc  Remove similarities at high-count positions (see Methods) \\
\   [Compute scores]   \\
\ccc  \textbf{for} each pair $(P,Q) \in P_s\times P_s$ \textbf{do}\\
\ccc  \qqa $S(P,Q) \gets 0$\\
\ccc  \textbf{for} each PPI $(P_1,P_2)\in P_i$ \textbf{do}   \\
\ccc  \qqa \textbf{for} each protein $Q_1$ and each similarity $(X_1,Y_1)$ in $(P_1,Q_1)$ \textbf{do}   \\
\ccc  \qqb \textbf{for} each protein $Q_2$ and each similarity $(X_2,Y_2)$ in $(P_2,Q_2)$ \textbf{do}   \\
\ccc  \qqc increase the score $S(Q_1,Q_2)$ by \eqref{eq_score_add}   \\
\   [Predict PPIs]\\
\ccc  sort the pairs in $P_s\times P_s$ decreasingly by score   \\
\ccc  \textbf{if} a threshold is provided  \\
\ccc  \qqa \textbf{then} output those with score above threshold   \\

\section{Declaration}


\subsection{Author's contributions}
L.I. proposed the problem, designed the SPRINT algorithm, computed the spaced seeds, and wrote the manuscript. Y.L. implemented the algorithm, contributed to its design and speed improvement, installed the competing programs, downloaded and processed the datasets and performed all tests.

\subsection{Funding}
L.I. has been partially supported by a Discovery Grant and a Research Tools and Instruments Grant from the Natural Sciences and Engineering Research Council of Canada (NSERC).

\subsection{Acknowledgements}
Evaluation has been performed on our Shadowfax cluster, which is part of the Shared Hierarchical Academic Research Computing Network (SHARCNET: http://www.sharcnet.ca) and Compute/Calcul Canada. We would like to thank the Sharcnet team, especially Edward Armstrong and Paul Preney, for helping us install a Debian image without which PPI-PK would not run at all.
L.I. has been partially supported by a Discovery Grant and a Research Tools and Instruments Grant from the Natural Sciences and Engineering Research Council of Canada (NSERC).

\subsection{Availability of data and materials}

Park and Marcotte's datasets are available from {\tt www.marcottelab.org/differentialGeneralization}. 

The source code of SPRINT is freely available from \texttt{https://github.com/lucian-ilie/SPRINT/}.

The UniProt protein sequences we used, precomputed similarities for these sequences, the datasets, and the top 1\% predicted PPIs for the entire human interactome can be found at \texttt{www.csd.uwo.ca/faculty/ilie/SPRINT/}.

\bibliographystyle{bmc-mathphys}
\bibliography{li_ilie}      

%
%
%
%

\end{document}